\documentclass[12pt,twoside]{article}   
\usepackage{array} 
\usepackage{multirow} 
\usepackage[super,sort,comma]{natbib}
\usepackage{amsmath} 
\usepackage{amssymb} 
\usepackage{tabularx} 
\newcolumntype{C}[1]{>{\centering\arraybackslash}p{#1}}
\newcolumntype{Y}{>{\centering\arraybackslash}X}
\usepackage{changes}

\usepackage{lipsum} 
\usepackage{geometry}
\usepackage{fancyhdr}		

\usepackage[section]{placeins}   %

\usepackage{graphicx}

\makeatletter \renewcommand\@biblabel[1]{$^{#1}$} \makeatother
\newcommand{\cen}[1]{\begin{center} #1 \end{center}}

\usepackage[mathlines]{lineno}

\usepackage{hyperref}
\hypersetup{ colorlinks,
    citecolor=blue,
    filecolor=blue,
    linkcolor=blue,
    urlcolor=blue
}

\usepackage{xcolor}

\definecolor{gray}{rgb}{0.6,0.6,0.6}
\definecolor{red}{rgb}{0.85,0,0}
\definecolor{green}{rgb}{0,0.85,0}
\definecolor{blue}{rgb}{0,0,0.85}
\definecolor{beige}{rgb}{0.92,0.87,0.78}
\usepackage[all]{hypcap}    

\begin{document}

\cen{\sf {\Large {\bfseries Actor Critic with Experience Replay-based automatic treatment planning for prostate cancer intensity modulated radiotherapy } \\  
\vspace*{10mm}
Md Mainul Abrar\textsuperscript{1}, Parvat Sapkota\textsuperscript{1}, Damon Sprouts\textsuperscript{1}, Xun Jia\textsuperscript{2, a}, Yujie Chi\textsuperscript{1, a}} \\
\textsuperscript{1}Department of Physics, The University of Texas at Arlington, Arlington, TX.\\
\textsuperscript{2}Department of Radiation Oncology and Molecular Radiation Sciences, Johns Hopkins University, Baltimore, MD.
\vspace{5mm}\\
Version typeset \today\\
}

\pagenumbering{roman}
\setcounter{page}{1}
\pagestyle{plain}
email: xunjia@jhu.edu, yujie.chi@uta.edu\\

\begin{abstract}
 
\noindent {\bf Background:} Achieving real-time treatment planning in intensity-modulated radiotherapy (IMRT) is challenging due to the complex interactions between radiation beams and the human body. The introduction of artificial intelligence (AI) has automated treatment planning, significantly improving efficiency. However, existing automatic treatment planning agents often rely on supervised or unsupervised AI models that require large datasets of high-quality patient data for training. Additionally, these networks are generally not universally applicable across patient cases from different institutions and can be vulnerable to adversarial attacks. Deep reinforcement learning (DRL), which mimics the trial-and-error process used by human planners, offers a promising new approach to address these challenges. \\ 
{\bf Purpose:} This work aims to develop a stochastic policy-based DRL agent for automatic treatment planning that facilitates effective training with limited datasets, universal applicability across diverse patient datasets, and robust performance under adversarial attacks. \\
{\bf Methods:} We employ an Actor-Critic with Experience Replay (ACER) architecture to develop the automatic treatment planning agent. This agent operates the treatment planning system (TPS) for inverse treatment planning by automatically tuning treatment planning parameters (TPPs). We use prostate cancer IMRT patient cases as our testbed, which includes one target and two organs at risk (OARs), along with 18 discrete TPP tuning actions. The network takes dose-volume histograms (DVHs) as input and outputs a policy for effective TPP tuning, accompanied by an evaluation function for that policy. Training utilizes DVHs from treatment plans generated by an in-house TPS under randomized TPPs for a single patient case, with validation conducted on two other independent cases. Both online asynchronous learning and offline, sample-efficient experience replay methods are employed to update the network parameters. After training, more than 300 initial treatment plans from three distinct datasets are used for testing. The ProKnow scoring system for prostate cancer IMRT, with a maximum score of 9, is used to evaluate plan quality. The robustness of the network is further assessed through adversarial attacks using the Fast Gradient Sign Method (FGSM).\\
{\bf Results:} Despite being trained on treatment plans from a single patient case, the network converges efficiently when validated on two independent cases. For testing performance, the mean $\pm$ standard deviation of the plan scores across all test cases before ACER-based treatment planning is $6.20 \pm 1.84$. After implementing ACER-based treatment planning, $93.09\%$ of the cases achieve a perfect score of 9, with only $6.12\%$ scoring between 8 and 9, and no cases being below 7. The corresponding mean $\pm$ standard deviation is $8.93 \pm 0.27$. This performance highlights the ACER agent's high generality across patient data from various sources. Further analysis indicates that the ACER agent effectively prioritizes leading reasonable TPP tuning actions over obviously unsuitable ones by several orders of magnitude, showing its efficacy. Additionally, results from FGSM attacks demonstrate that the ACER-based agent remains comparatively robust against various levels of perturbation.\\
{\bf Conclusions:} We successfully trained a DRL agent using the ACER technique for high-quality treatment planning in prostate cancer IMRT. It achieves high generality across diverse patient datasets and exhibits high robustness against adversarial attacks. \\

\end{abstract}

\newpage     



\setlength{\baselineskip}{0.7cm}      

\pagenumbering{arabic}
\setcounter{page}{1}
\pagestyle{fancy}
\section{Introduction}
Real-time treatment planning represents a challenging problem in intensity modulated radiotherapy (IMRT) for cancer treatment. The difficulty mainly arises from the complex energy deposition properties of the radiation beam within the patient body, which often requires a careful balance between target dose coverage and normal tissue sparing. This balance can be achieved via inverse treatment planning optimization, which involves setting dose constraints for each organ and target, along with weighting factors to prioritize conflicting dose requirements. However, the optimal value set of these treatment planning parameters (TPPs) to achieve clinical acceptable plan are often initialization condition dependent, such as the patient geometries, fluencemap and TPP initialization, etc. \cite{webb2003physical}. In state-of-the-art radiotherapy clinics, it requires repeatedly adjusting the TPPs by human planners to refine the inverse treatment planning optimization, which is time-consuming and poses challenges for real-time planning.

In the era of artificial intelligence (AI), significant efforts have been made to address challenges in treatment planning. One approach aims to shorten the trial-and-error process by generating a relatively "good" initial plan using machine learning, allowing human planners to refine and optimize the plan more efficiently \cite{li2020automatic}. For instance, Li \textit{et al.} \cite{li2020automatic} successfully employed architectures like ResNet \cite{7780459} and DenseNet \cite{8099726} to create high-quality starting plans, thereby accelerating the overall treatment planning process. Another approach bypasses the trial-and-error process entirely by using machine learning to directly predict desired 3D dose distributions and corresponding 2D fluence map intensities based on patient image and contour data \cite{vandewinckele2022treatment, lempart2021volumetric}. For example, Vandewinckele \textit{et al.} \cite{vandewinckele2022treatment} utilized a U-Net-based convolutional neural network (CNN) \cite{ronneberger2015u} to predict 3D dose distributions from CT scans and the contours of targets and organs. They subsequently applied another U-Net-based CNN to predict the 2D fluence map from the 3D dose, with or without patient image and contour data, for lung cancer IMRT cases.

While these AI techniques have significantly reduced the time required for treatment planning, they often depend on large databases of high-quality patient data for training due to their supervised or unsupervised learning nature. Additionally, the trained networks are frequently not universally applicable across different institutions due to data heterogeneity \cite{kandalan2020dose}, creating barriers to their widespread adoption. Furthermore, these developments typically lack adversarial attack testing, despite vulnerabilities being identified in various supervised learning algorithms used in medical applications \cite{finlayson2019adversarial}.

Reinforcement learning (RL) \cite{sutton2018reinforcement}, which mimics the trial-and-error learning process used by humans to achieve their goals, offers a new angle to accelerate treatment planning. Unlike other methods, the trial-and-error process in RL generates substantial new data samples that the algorithm can learn from, thereby reducing the dependence on large initial datasets. Recently, the application of deep neural network-based Q-learning, specifically Deep Q-Networks (DQN) \cite{mnih2015human}, has shown promising results in automating treatment planning \cite{shen2019intelligent, shen2020operating, shen2021improving, shen2021hierarchical, sprouts2022development, gao2023implementation}. For example, Shen \textit{et al.} \cite{shen2019intelligent} utilized DQN to develop a network that observed dose volume histograms (DVHs) and output actions to adjust organ weighting factors in inverse treatment planning for high-dose-rate brachytherapy in cervical cancer. The same research group later demonstrated the feasibility of this approach for automatic tuning of TPPs in external beam IMRT for prostate cancer \cite{shen2020operating}. Additionally, Sprouts \textit{et al.} \cite{sprouts2022development} extended the DQN-based virtual treatment planner (VTP) to adjust TPPs compatible with commercial TPS systems, achieving effective treatment planning for prostate IMRT. 

Despite these promising advancements, the DQN network has inherent limitations that complicate its application to complex, clinically relevant treatment planning scenarios. In clinical settings, human planners often adjust numerous TPPs for different targets and organs at risk (OARs), resulting in a vast state-action space that the RL algorithm must explore. In such cases, finding the optimal action concerning the Q function can be costly due to challenges like overestimation of the Q value function and difficulties in balancing exploration and exploitation \cite{zhu2021overview}. To partially address these issues, Shen \textit{et al.} introduced a knowledge-guided network training strategy \cite{shen2021improving} and a hierarchical approach \cite{shen2021hierarchical} within the DQN framework, demonstrating some success in prostate Stereotactic Body Radiation Therapy (SBRT) automatic treatment planning \cite{gao2023implementation}. However, not all challenges were resolved. Additionally, continuous TPP tuning leads to a continuous action space, where DQN's effectiveness diminishes. Like other deep neural networks, DQNs are also vulnerable to adversarial attacks \cite{huang2017adversarial}, raising concerns about their robustness in clinical applications. 

To tackle these challenges, we propose using a new RL approach: the Actor-Critic Experience Replay (ACER) \cite{wang2016sample} method to automate the treatment planning process. ACER builds on the advanced actor-critic algorithm (A3C) \cite{mnih2016asynchronous}, enhancing data sampling efficiency through experience replay \cite{lin1992self}. In this framework, the actor functions as the policy network, aiming to maximize returns, while the critic assesses the quality of the actor's decisions. This setup inherently facilitates exploration and exploitation, and the policy gradient method allows effective exploration of both discrete and continuous action spaces. Previous studies indicate that A3C can be more resistant to adversarial attacks than DQN \cite{huang2017adversarial}. Based on these observations, we propose applying ACER to develop a training-efficient, robust, and scalable agent for automatic treatment planning applications.

In the following sections, we will detail our implementation of the ACER algorithm for automating the TPP tuning process in inverse treatment planning, using prostate IMRT as a test case. Followed by it is the evaluation of its performance across different datasets, including its vulnerability to adversarial attacks.

\section{Methods and Materials}
\subsection{Overall Architecture}
The overall architecture of the ACER based automatic treatment planning system is similiar to that of the DQN-based system \cite{sprouts2022development}, as shown in Figure 1. The process begins with the random initialization of the TPPs, which are then input into the TPS system for inverse treatment planning. The quality of the resulting treatment plan is evaluated, and if it does not meet the desired standards, both the plan and the TPPs are fed into the ACER-based VTP system for TPP tuning. With the updated TPPs, the TPS performs inverse treatment planning optimization again. This iterative process continues until the plan quality meets the required standards or the maximum number of TPP tuning iterations is reached.\\[0.25 cm]
\begin{figure}[ht] 
    \centering
    \includegraphics[width=0.9\textwidth]{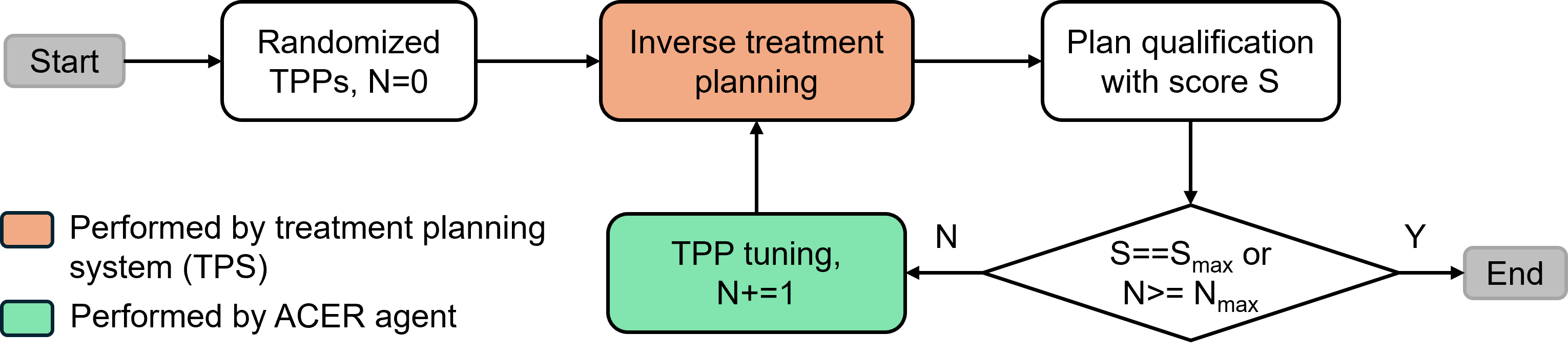} 
    \caption{The workflow of the actor critic with experience replay (ACER)-based automatic treatment planning process. Here, 'TPP' represents treatment planning parameter.}
    \label{fig:Workflow_algorithm}
\end{figure}

In the following subsections, we will give the design details for the in-house TPS, the ACER-based VTP system, the testbed, the plan evaluation system, and the system performance test.

\subsection{In-house Treatment Planning System (TPS)}

We developed an in-house dose-volume constraint TPS \cite{sprouts2022development} following the documentation of Varian's Eclipse \cite{varian2014eclipse}. For a treatment planning containing a single target and $N$ OARs, the objective function for the fluencemap optimization can be formulated as:
\begin{multline}
    min \frac{1}{2}||Mx-d_p||^{2}_{-}+ \frac{\lambda}{2}||(Mx-td_p)_{V_{\mathrm{PTV}}}||^{2}_{+}+\sum_i^N \frac{\lambda_i}{2}||(M_ix-t_{i}d_p)_{V_i}||^{2}_{+}, \\
    \textrm{s.t. } x \geq 0, D_{95\%}(Mx) = d_p.
\end{multline}
In this equation, $|\cdot|^2_-$ and $|\cdot|^2_+$ represent the standard $l_2$ norms, which compute only the negative and positive elements, corresponding to under-dose and over-dose constraints, respectively. The under-dose constraint is further reinforced that at least $95\%$ of the PTV volume receives the prescription dose. The relative importance of the respective terms is adjusted by the weighting factors $\lambda$ and $\lambda_i$. Regarding the other variables in the equation, $M$ represents the dose deposition matrix, $x$ is the beamlet vector, and $d_p$ is the prescription dose. $td_p$ and $V_{\mathrm{PTV}}$ are upper threshold dose and the percentage of the PTV volume considered for over-dose constraints, respectively. Similarly, $M_i$, $V_i$, and $t_i$ are the corresponding variables used for the over-dose constraints on the $i$th OAR. Voxels included in $V_{\mathrm{PTV}}$ and $V_i$ always receive higher dose than those not selected. 

In summary, the free TPPs to be tuned in this in-house TPS are $\lambda$, $V_{\mathrm{PTV}}$, $t$, $\lambda_i$, $V_i$, and $t^d_i$ (where $i=1, 2, ..., N$), totaling $3(N+1)$ parameters. Given a set of TPPs, the optimization problem can be solved using the alternating direction method of multipliers (ADMM). 

\subsection{Actor Critic with Experience Replay (ACER)-based Virtual Treatment Planner (VTP) System}
\subsubsection{The working principle of ACER}
ACER is a type of deep reinforcement learning that integrates a deep neural network with actor-critic learning while leveraging the experience replay \cite{wang2016sample}. This approach has shown superior performance in challenging environments, including the Atari57 game collection \cite{wang2016sample}. 

In actor-critic learning, an actor agent generates decisions while a critic agent evaluates those decisions in the context of a sequential decision-making problem. This process involves a dynamic environment represented by a series of states $s\in\mathcal{S}$ associated with a series of possible actions $a\in\mathcal{A}$. After taking an action $a$ in state $s$, the state transitions into the next state $s'$ with a probability $Pr\{s'|s, a\}$, yielding a stepwise reward $r\in \mathcal{R}$. The optimal decision made by the actor agent, or the optimal policy $\pi$, can be defined as choosing those series of actions $a$ to maximize the accumulated reward for states $s$ over time $t\in(0,1,2...)$. The corresponding objective function is \cite{sutton1999policy}:
\begin{equation}
    J(\pi)= \sum_{s}\textrm{lim}_{t\to\infty}Pr\{s_t=s|s_0,\pi\}\sum_{a}\pi(a|s) \mathcal{R}^{a}_s.
\end{equation}

Representing the policy $\pi$ by a network that has parameters $\theta$, the policy parameters $\theta$ can be optimized using the policy gradient approach governed by the policy gradient theorem \cite{sutton1999policy} as:
\begin{align}
\begin{split}
        g&=\nabla_{\theta}J(\pi_{\theta})\\
        &= \sum_{s}\textrm{lim}_{t\to\infty}Pr\{s_t=s|s_0,\pi\}\sum_{a}\ \nabla_{\theta}\pi_{\theta}(a|s) Q^{\pi_{\theta}}(a,s)\\
    &=\mathbb{E}_{\theta}[Q^{\pi_{\theta}}(a,s)\nabla_{\theta}\mathrm{log}\pi_{\theta}(a|s)].
    \end{split}
\end{align}
Here, $Q^{\pi}(a,s)$ is the state-action value function, serving as an effective evaluation of the policy performance. 

In actor-critic learning, $Q^{\pi}(a,s)$ is estimated by the critic agent. When selecting an action $a$ in a state $s$ at step $t$, the critic estimates $Q^{\pi}(a_t,s_t)$ as the expected cumulative reward 
\begin{equation}
\mathbb{E}_{s_{t+1}:\infty, a_{t+1}:\infty}\left(\sum_{i\geq 0}\gamma^i r_{t+i}|a_t=a,s_t=s\right) 
\end{equation}
following policy $\pi$. Here, $\gamma \in [0, 1)$ is the discount factor for future rewards.

In ACER, the actor and critic agents are integrated into a deep neural network with 'two heads'. One head gives the policy $\pi_{\theta}(a_t,s_t)$ with network parameters $\theta$, while the other outputs the estimate $Q_{\theta_c}(a_t,s_t)$ with network parameters $\theta_c$. Particularly, ACER designs the policy network to contain two parts, the distribution $f$, and the statistics $\phi_{\theta}(s)$ of $f$, thus the policy can be fully represented as $\pi(\cdot|s)=f(\cdot|\phi_{\theta}(s))$. 

To solve the intrinsic instability that arises from combining online RL with deep neural network \cite{mnih2016asynchronous}, ACER employs a hybrid online and offline training strategy to optimize the network performance. Specifically, it adopts the online asynchronous updating of $\theta$ and $\theta_c$ by launching parallel network learners that interact with different instances of the environment \cite{mnih2016asynchronous}. This technique de-correlates the agents’ data, thereby stabilizing the learning process. Additionally, it implements the offline experience replay strategy \cite{lin1992self}, allowing the agent to learn from memory of experiences. Trajectories can be retrieved from the memory and weighted by importance sampling, promoting both learning stability and data efficiency.

Particular to the training of the critic network, ACER constructs the target $Q$ with retrace estimator \cite{munos2016safe} as 
\begin{align}
\begin{split}
        Q^{\mathrm{tar}}(a_t,s_t)&=Q^{\mathrm{ret}}(a_t,s_t)\\
        &= r_{t}+\gamma\bar{\rho}_{t+1}[Q^{\mathrm{ret}}(a_{t+1},s_{t+1})-Q_{\theta_c}(a_{t+1},s_{t+1})]+\gamma V_{\theta_c}(s_{t+1}).
\end{split}        
\end{align}
Here, $\bar{\rho}=min(c, \rho)$ is the truncated importance weight used in experience replay, where $c$ is a constant and $\rho$ is the importance weight. During online updating, $\bar{\rho}=1$. The value function $V_{\theta_c}(s)$ is derived from the critic's $Q$ estimator as $V_{\theta_c}(s)=\Sigma_{a}Q_{\theta_c}(a|s)f(a|\phi(s))$. $Q^{\mathrm{ret}}$ has been shown to have low variance and to converge effectively. With it, the critic network parameter $\theta_c$ is then updated as $d\theta_c=d\theta_c+\nabla_{\theta_c}(Q^{\mathrm{ret}}(a,s)-Q_{\theta_c}(a,s))^2$. 

The low-variance $Q^{\mathrm{ret}}$  is also employed to stabilize online policy updates by replacing $Q^{\pi_\theta}$ term as $Q^{\mathrm{ret}}-V$ in Eq. (3). ACER further incorporates truncated importance sampling in experience replay with bias correction to enhance data efficiency while avoiding excessive bias in policy updates, which yields the offline policy gradient $g$ with respect to $\phi$ as follows:
\begin{multline}
    g^{\mathrm{ACER}}_t = \bar{\rho}_t \nabla_{\phi_{\theta}(s_t)} \log{f(a_t|\phi_\theta (s_t))}[Q^{\mathrm{ret}}(a_t,s_t)-V_t(s_t)]\\
    + \underset{a\sim \pi}{\mathbb{E}}\left( [\frac{\rho_t (a)-c}{\rho_t (a)}]_+ \nabla_{\phi_{\theta}(s_t)}\log{f(a_t|\phi_\theta (s_t))} [Q_{\theta_{c}}(a, s_t)-V_{\theta_{c}}(s_t)]\right).
\end{multline}

Finally, to limit the per-step changes to the policy and achieve stability, ACER provides the option to utilize a modified version of Trust Region Policy Optimization (TRPO) \cite{schulman2015trust}, ensuring that the updated policy does not deviate significantly from the average policy network $\phi_a$. Specifically, it restricts the policy network parameter $\theta$ updating at step $t$ as \begin{equation}
 d\theta=d\theta+\frac{\partial \phi_{\theta}(s_t)}{\partial \theta}\left(g^{\mathrm{ACER}}_t-\mathrm{max}\{0,\frac{k^Tg^{\mathrm{ACER}}_t-\delta}{||k||^2_2}\}k\right).
 \end{equation}
 Here, $k=\nabla_{\phi_{\theta}(s_t)}D_{KL}[f(\cdot|\phi_{\theta_a}(s_t))||f(\cdot|\phi_{\theta}(s_t))]$ is the linear Kullback–Leibler divergence ($D_{KL}$) and $\delta$ is the predefined divergence constraint. In this situation of disabled TRPO updating, the policy network parameter is updated as
\begin{equation}
d\theta=d\theta+\frac{\partial \phi_{\theta}(s_t)}{\partial \theta}g^{\mathrm{ACER}}
\end{equation}

In practice, entropy regularization term $\beta\nabla_{\theta}H(\pi(s,\theta))$ \cite{mnih2016asynchronous} can also be used to boost online and offline policy update performance, with $\beta$ the weighting parameter. This term improves exploration by discouraging premature convergence to sub-optimal deterministic policies.

With variables $d\theta_c$ and $d\theta$ computed, the network parameters $\theta_c$ and $\theta$ are updated with RMSprop algorithm \cite{tieleman2012lecture}. Under active TRPO updating, the parameters $\theta_a$ for the average policy network $\phi_a$ is updated as $\theta_a=\alpha \theta_a+(1-\alpha)\theta$, with $\alpha$ the average model decay rate.

\subsubsection{The establishment of the VTP system upon ACER architecture}
We apply the ACER architecture to develop the VTP system for automatic treatment planning as follows. We define an input state $s$ as discrete points taken from the DVH curves of a treatment plan. For a treatment plan containing $M$ target and OARs, the input state $s$ has a dimension of $m\times M$, where $m$ represents the number of discrete points on each DVH curve. The action space consists of tuning strategies for the TPPs. In this initial approach, we allow each TPP to have two tuning strategies: increasing or decreasing by a predefined amount. With $M$ target and OARs, there are $3M$ TPPs to tune, resulting in an action space of length $6M$. This defines the dimensions of both the policy distribution and the $Q$-value function space, each with length $6M$. 
 
Once a TPP tuning action $a$ is predicted for treatment plan $s$ at time $t$, the in-house TPS performs inverse treatment planning to generate a new plan $s_{t+1}$. For each state-action pair $(s_t,a_t)$, the immediate reward $r$ is calculated as the difference in plan quality between the new plan and the current plan, that is $r_t=\psi(s_{t+1})-\psi(s_t)$, where $\psi(s)$ represents the quality evaluation of the plan $s$. The total reward is computed as $R=\sum^{t_\mathrm{epi}-1}_{i=t}\gamma^{i-t}r_i$, reflecting the accumulated return across all future plans in that episode.

Training of the ACER-based VTP follows the standard ACER training process. We launch $N_a$ parallel online training agents, each with distinct input treatment plans. The agents asynchronously update the network policy upon completing an episode with a maximum step length of $t_{\mathrm{epi}}$. After each episode, each agent is restarted with a different input treatment plan. During the process, each agent stores its episodic trajectories in a replay buffer with a storage length of $t_{\mathrm{rep}}$ steps. Once $t_s$ steps have been accumulated, offline policy training examines the experience pool using a batch size of $t_B$ steps and begins to update the policy network at a frequency $p-1$ times higher than that of the online updates. To consistently monitor training performance, we evaluate the process every $t_{\mathrm{eval}}$ steps using independent evaluation patient cases. The training continues until a maximum of $T_{\mathrm{max}}$ steps is reached.

\subsubsection{Testbed}\label{testbed}
In line with our previous development efforts \cite{sprouts2022development}, this paper continues to use IMRT treatment planning for prostate cancer as a testbed to test the proposed ACER-based VTP system. We consider a scenario involving one target (the prostate) and two OARs (the bladder and the rectum), leading to the optimization of three DVH curves. Each curve is represented by 100 discrete points, resulting in a total of 300 floating values as the input to the ACER agent. According to Eq. (1), we have 9 TPPs to tune, creating an action space of 18, as shown in Table \ref{table: actions}. The specific increment and decrement amplitudes are determined based on experience and are not expected to significantly affect the overall convergence of the treatment planning process. 
\begin{table}[h!]
\caption{The actions to tune the treatment planning parameters (TPPs) in step $t$ based on the TPP values in step $t-1$. Here, 'OAR' represents both rectum and bladder.}
\label{table: actions}
\centering
\renewcommand{\arraystretch}{1.3}
\begin{tabular}{m{2 cm}m{2.7 cm}m{2.5 cm}m{2.5 cm}m{2.5 cm}m{2.0 cm}}
  \hline
  actions & $\lambda^{t}_{\mathrm{PTV,OAR}}$ & $t^{t}_{\mathrm{PTV}}$ & $t^{t}_{\mathrm{OAR}}$ & $V^{t}_{\mathrm{PTV}}$ & $V^{t}_{\mathrm{OAR}}$ \\
  \hline
  increase & $1.65\lambda^{t-1}_\mathrm{PTV,OAR}$ & $\mathrm{min}(1.2,$ $ 1.01t^{t-1}_\mathrm{PTV})$ & {$\mathrm{min}(1, $ $  1.25t^{t-1}_\mathrm{OAR})$} & $\mathrm{min}(0.3, $ $  1.25V^{t-1}_\mathrm{PTV})$ & $\mathrm{min}(1, $ $  1.25V^{t-1}_\mathrm{OAR})$ \\
  decrease & $0.6\lambda^{t-1}_\mathrm{PTV,OAR}$ & $\mathrm{max}(1, $ $  0.91t^{t-1}_\mathrm{PTV})$ & $0.6t^{t-1}_\mathrm{OAR}$ & $0.8V^{t-1}_\mathrm{PTV}$ & $0.8V^{t-1}_\mathrm{OAR}$ \\
  \hline
\end{tabular}
\end{table}

We use the ProKnow scoring system (ProKnow Systems, Sanford, FL, USA) for prostate cancer IMRT plans to estimate the plan quality $\psi(s)$. Nine original scoring criteria from the ProKnow system relevant to our testbed are shown in Table \ref{table:TPPtuning}. In practice, we slightly adjust these criteria \cite{shen2021improving} and apply them to compute $\psi(s)$ for each plan $s$. With equal weighting on each criterion, the score ranges from 0 to 9. 

\begin{figure}[!htbp] 
    \centering
    \includegraphics[width=1\textwidth]{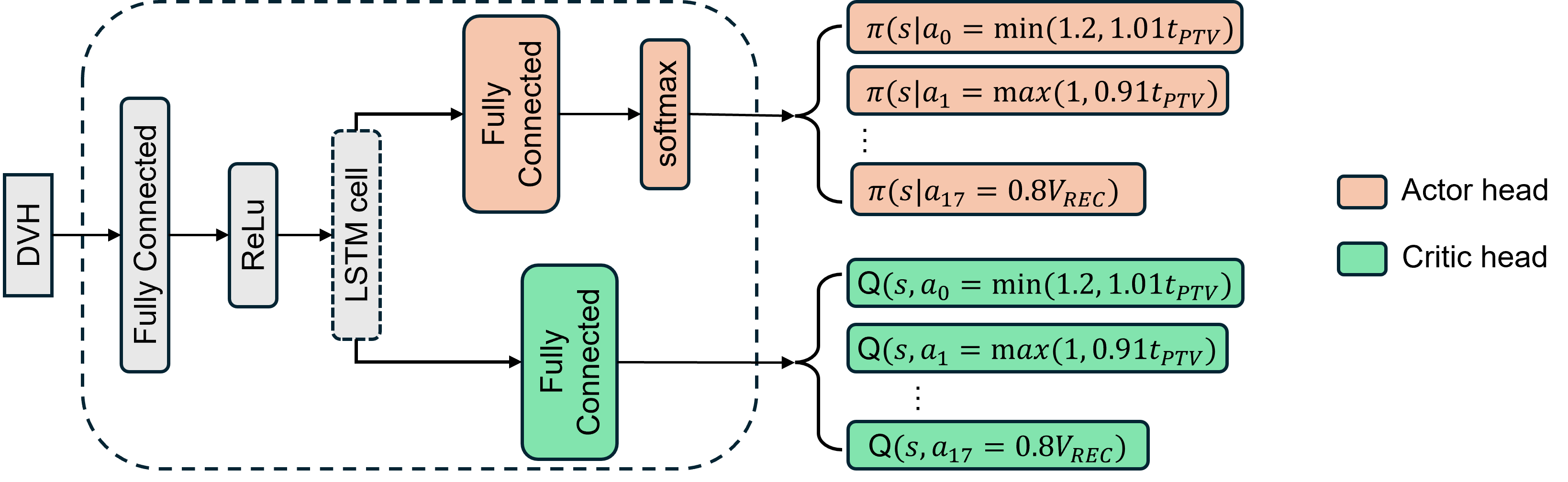} 
    \caption{The deep neural network featuring a 'two-head' structure designed for the actor critic with experience replay (ACER)-based virtual treatment planner (VTP). It takes the dose volume histogram (DVH) of the current plan as input. It outputs a treatment planning parameter (TPP) tuning strategy across 18 actions in one head, and produces the corresponding Q-value in the other head.}
    \label{fig:ACERstructure}
\end{figure}

\begin{table}[h!]
    \caption{The nine criteria originated from the planIQ scoring system relevant to this study.}
    \label{table:TPPtuning}
    \centering
    \renewcommand{\arraystretch}{1.3}
    \begin{tabularx}{\textwidth}{c*{9}{Y}}
        \hline
        \multicolumn{4}{c|}{Bladder} & \multicolumn{4}{c|}{Rectum} & {PTV} \\ \cline{1-9}
        $V_{100.6\%}$ & $V_{94.3\%}$ & $V_{88.1\%}$ & $V_{81.2\%}$ & $V_{94.3\%}$ & $V_{88.1\%}$ & $V_{81.2\%}$ & $V_{75.5\%}$ & $D_{0.03cc}$ \\ \hline
        $<20\%$ & $<30\%$ & $<40\%$ & $<55\%$ & $<20\%$ & $<30\%$ & $<40\%$ & $<55\%$ & $<109.6\%$ \\ \hline
        {}\\
    \end{tabularx}
\end{table}
For this testbed, the specific network architecture of the ACER-based VTP system is shown in Figure \ref{fig:ACERstructure}. The input DVH data pass through a fully connected layer with a hidden size of 32, followed by a Rectified Linear Unit (ReLU) activation layer. They then enter a long short-term memory (LSTM) layer with a hidden size of 32. Afterward, the data are split: one path goes through a fully connected layer followed by a softmax layer, which outputs the policy, while the other path flows into a separate fully connected layer to produce the $Q$ value function. The implementation of the network relies on the pytorch and Open AI Gym libraries. The learnable parameters are distributed across the three fully connected layers and the LSTM cell, totaling around 20,000 parameters. \\[0.25cm]

We use three independent datasets of prostate cancer IMRT cases for network training, validation and testing. The first dataset contains 52 independent patient treatment plans, as described in our previous development \cite{sprouts2022development}. The second dataset is the Common Optimization for Radiation Therapy (CORT) dataset from the Mass General Radiation Oncology Physics Division, which is publicly available and contains one independent patient case \cite{craft2014shared}. The third dataset is The Radiotherapy Optimization Test Set (TROTS), which includes 30 independent patient cases \cite{breedveld2017data}. 

It is worth noting that for each dataset, by randomly initializing the TPP values, we can generate numerous independent initial treatment plans to form different state-action-reward trajectories. In this study, to demonstrate the network training efficacy upon limited patient cases, we use one patient case in dataset 1 for training, two independent patient cases in dataset 1 for validation, and all remaining cases across the three datasets for testing.

\subsection{Adversarial Attack}
After completing the training of the ACER agent, except for testing its performance on TPP tuning decisions, we also evaluate its robustness against adversarial attacks. Adversarial attacks are malicious attempts to manipulate machine learning models into making incorrect predictions or decisions \cite{szegedy2013intriguing, biggio2018wild}. Given the potential clinical application of automatic treatment planning agents in the future, it is crucial to train the ACER agent to be robust against such attacks \cite{finlayson2019adversarial}.  

We assume the adversary has access to the trained policy network, allowing it to fool the network by perturbing the input state in a way that exploits the policy's sensitivity. Specifically, we use the Fast Gradient Sign Method (FGSM) \cite{goodfellow2014explaining} to compute the state perturbation $\eta$, which is the gradient of loss function $J$ with respect to the state $s$: 
\begin{equation}
    \eta = \epsilon \mathrm{sign}(\nabla_s J(\theta, s,y )).
\end{equation}
Here, $\epsilon$ is a constant, serving as the upper limit for the element-wise perturbation, i.e., $|\eta|_\infty \leq \epsilon$. $y$ is the distribution over all possible actions. With this, the perturbed state becomes $s' = s + \eta$.

We apply the FGSM-based attack to both the ACER agent trained in this work and the DQN agent from our previous study \cite{sprouts2022development}, and compare their robustness to the attack. In the ACER agent, $y$ is the stochastic policy $\pi$. $J(\theta, s, y)$ is represented as the cross entropy loss between $y$ and the distribution that places all weight on the highest-weighted action $a_j$ in $y$. Specifically 
\begin{equation}
    J(\theta,s,\pi) = -\frac{1}{18}\left(\mathrm{log}(\pi^\theta(a_j|s))+\sum_{i\neq j}\mathrm{log}(1-\pi^\theta(a_i|s))\right).
\end{equation}

In the DQN agent, the policy determined by the Q value function is deterministic, which causes the problem that the gradient of $J(\theta, s, y)$ is almost zero for all input states. To solve it, we define $y$ as the softmax of the $Q$ value function \cite{huang2017adversarial}. We set $\epsilon$ to be values of 0.001, 0.01, and 0.1, apply the corresponding attacks and record their perturbations on action priorities and the next-step treatment plan qualities.

\section{Results}
\subsection{Training results}
\begin{table}[h!]
\caption{The hyperparameters and their values used to train the actor critic with experience replay (ACER)-based virtual treatment planner (VTP) for intensity modulated radiotherapy (IMRT) treatment planning of prostate cancer.}
\label{table: hyperparameters}
\centering
\renewcommand{\arraystretch}{1.3}
\begin{tabular}{>{\raggedright\arraybackslash}m{3.5cm}>{\raggedright\arraybackslash}m{3.5 cm}>{\raggedright\arraybackslash}m{8cm}}
\hline
\textbf{Hyperparameter} & \textbf{Value} & \textbf{Description} \\
\hline
 $N_a$ & 3 & Number of asynchronous training agents \\
$T_{\mathrm{max}}$ & 250000 & Number of total training steps \\
$t_{\mathrm{epi}}$ & 20 & Maximum length of an episode\\
$t_{\mathrm{rep}}$ & 100000 & Storage capacity of experience replay memory \\
$t_{s}$ & 2000 & Number of accumulated transitions before starting off-policy training \\
$t_{B}$ & 16 & Off-policy batch size \\
$c$ & 10 &  Importance-weight truncation in experience replay\\
$p$ & 4 & Ratio of off-policy to on-policy updates \\
$t_{\mathrm{eval}}$  & 500 & Interval between two adjacent evaluations \\
$\gamma$ & 0.99 & Discount factor \\
$\beta$ & 0.001 & Weighting for entropy regularization\\
 $\alpha$ & disabled & Decay rate for the average policy model\\
 $\delta$ & disabled & Trust region threshold value \\
\hline
\end{tabular}
\end{table}
The hyperparameter values used to configure the ACER-based VTP for prostate cancer IMRT treatment planning are listed in Table \ref{table: hyperparameters}. We utilize 3 CPU cores for online asynchronous training, aiming to complete the entire training within 250,000 steps. To encourage early convergence in the treatment planning process, we limit the length of each episode to 20 steps. Offline training begins with the storage of 2,000 steps of experiences, with a total maximum experience storage capacity of 100,000 steps. The TRPO-based updating method has been found to be computationally expensive; therefore, we disable it in this study to simplify the network. The values for the remaining parameters in Table \ref{table: hyperparameters} are consistent with those used for training the ACER agent in the Atari57 game set \cite{wang2016sample}. The entire training process takes approximately 7 days on an Intel(R) Core(TM) i7-6850K CPU @ 3.60GHz.

\begin{figure}[!htbp] 
    \centering
    \includegraphics[width=0.8\textwidth]{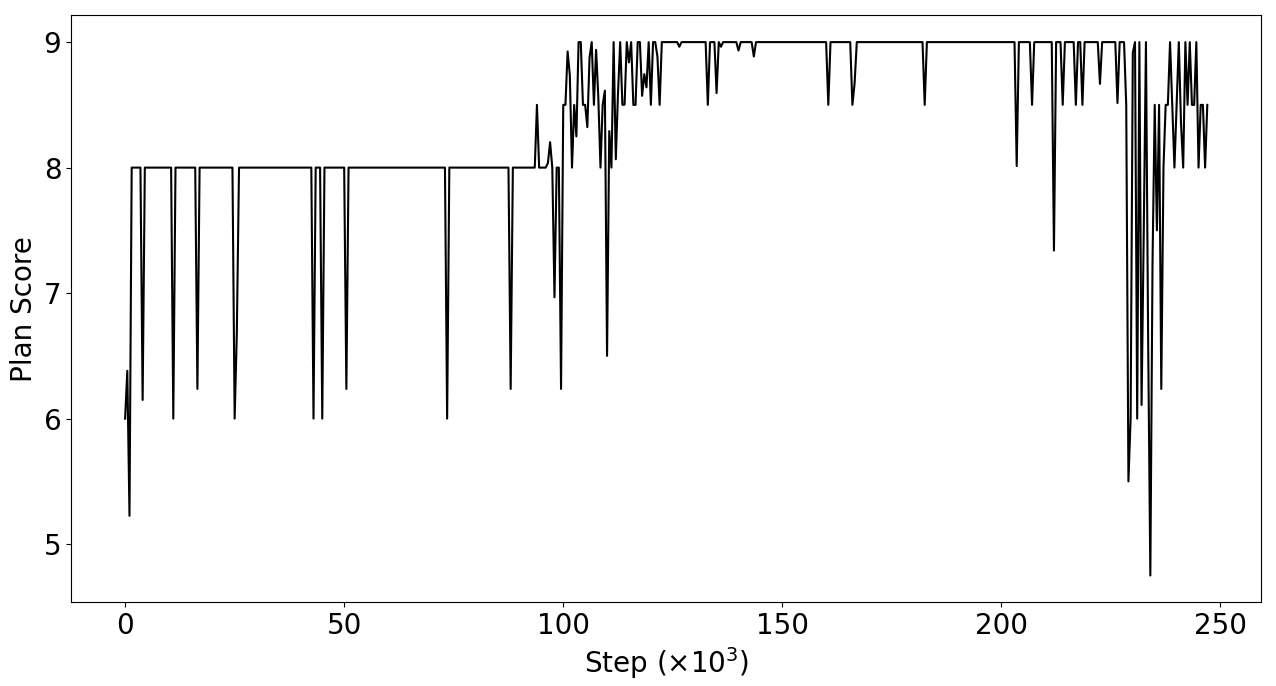} 
    \caption{The convergence map of the actor critic with experience replay (ACER)-based virtual treatment planner (VTP) training process, evaluated based on the average plan score for the validation patient cases.}
    \label{fig:ConvergenceMap}
\end{figure}

The convergence map of the agent training process, evaluated based on the average plan score for the validation patient cases, is shown in Figure \ref{fig:ConvergenceMap}. As illustrated, the plan score gradually approaches the maximum value of 9 as training progresses, with reduced fluctuations until approximately 200,000 steps. Beyond this point, performance becomes unstable, exhibiting large fluctuations, which we interpret as overfitting to the training cases. Therefore, we select the policy obtained at an earlier convergence point, around step 120,500, for testing.\\[0.25 cm]

\subsection{Testing results}
To demonstrate the efficacy of ACER-guided automatic treatment planning, we first test the network using 49 patient cases independent of the training and validation cases from dataset 1 and compare the results with DQN-based treatment planning\cite{sprouts2022development}. The TPPs are initialized with trivial values (all set to 1, except for $V_\mathrm{PTV}=0.1$), as done in our previous work\cite{sprouts2022development}. \\[0.25 cm]
\begin{figure}[htbp] 
    \centering
    \includegraphics[width=0.9\textwidth]{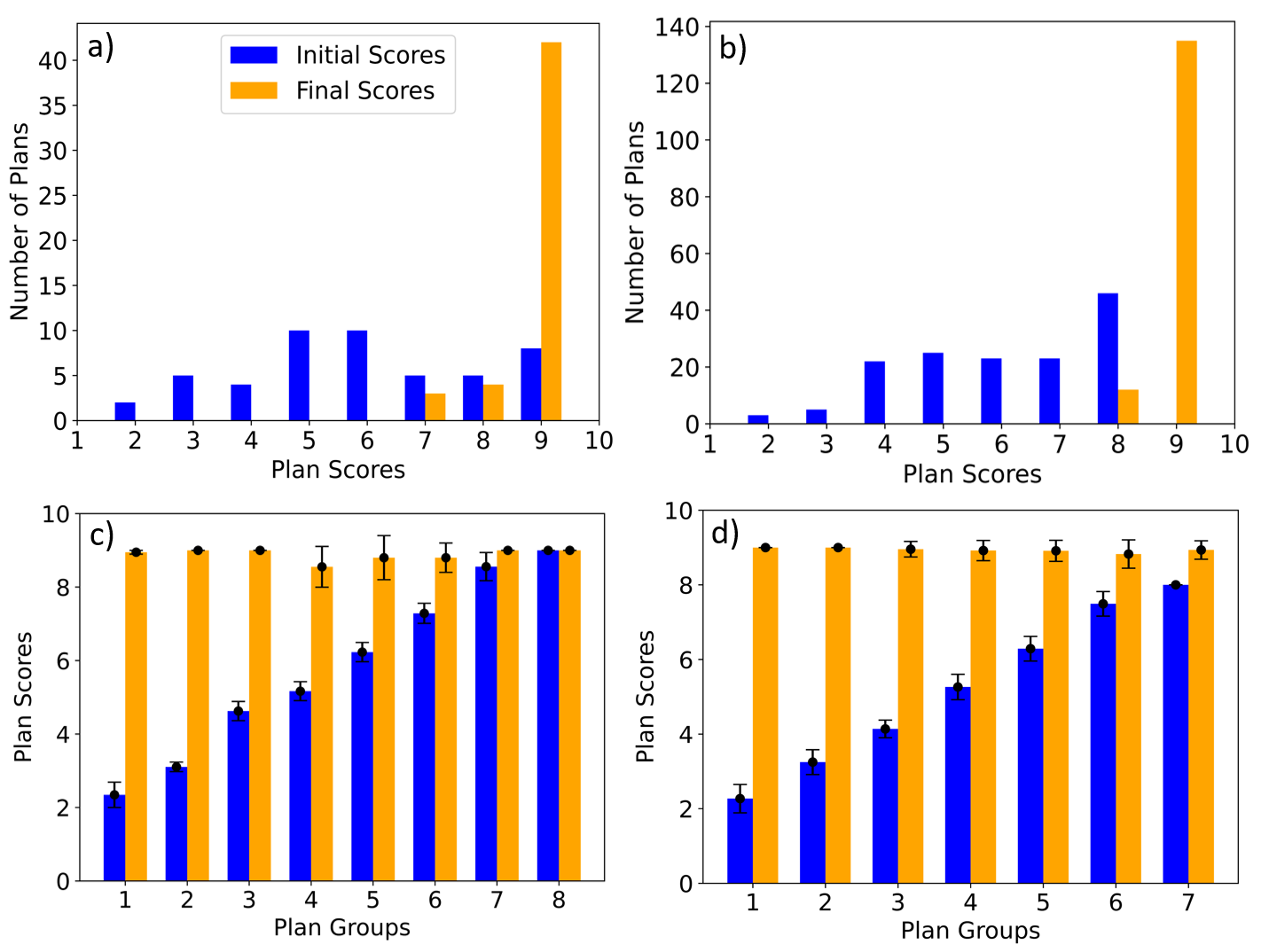} 
    \caption{(a)-(b) The plan score distributions for 49 test cases generated under trivial treatment planning parameter (TPP) settings and 147 test cases generated under random TPP settings from 49 patient cases in dataset 1, respectively, before and after actor critic with experience replay (ACER)-guided treatment planning. The histogram width is set to 1. (c)-(d) The mean and standard deviation of the plan score distributions before and after ACER-based treatment planning for the cases shown in (a) and (b), respectively. The groups in (c) and (d) correspond one-to-one with the histogram distributions in (a) and (b). ACER-based treatment planning significantly improves plan quality, achieving a mean score close to 9 across all plan groups.}
    \label{fig:dataset1}
\end{figure}

The results are shown in Figure \ref{fig:dataset1} (a) and (c). In Figure \ref{fig:dataset1} (a), the patient cases are grouped into 8 categories based on their plan scores (group index $i\in (1,2,...,8)$ includes cases with plan scores in the range of $[i,i+1)$). Before ACER-guided treatment planning, plan scores are distributed broadly from 2 to 9, with a mean score and standard deviation (std.) of $6.20 \pm 2.01$. After ACER-guided treatment planning, 42 out of 49 cases achieve a full score of 9, 1 case reaches 8.9, 3 cases score 8, and 3 cases score between 7 and 8. The corresponding mean and std. are $8.85 \pm 0.43$. In comparison, DQN-guided treatment planning improves the plan score from $6.18 \pm 1.75$ to $8.14 \pm 1.27$ for 50 patient cases from the same patient dataset\cite{sprouts2022development}. 

In Figure \ref{fig:dataset1} (c), the same 49 patient cases are divided into 8 groups based on their initial plan scores. The mean and standard deviation of the plan scores for each group, both before and after ACER-based treatment planning, are plotted. It is evident that after ACER-guided treatment planning, the plan scores are uniformly improved, approaching 9 across all patient groups, including those with very low initial scores below 3 (patient group 1). In contrast, DQN-based treatment planning shows that some patients with low initial scores could not be efficiently improved (as depicted in Figure 5(a) in Sprouts \textit{et. al} \cite{sprouts2022development}).

\begin{figure}[h!] 
    \centering
    \includegraphics[width=0.9\textwidth]{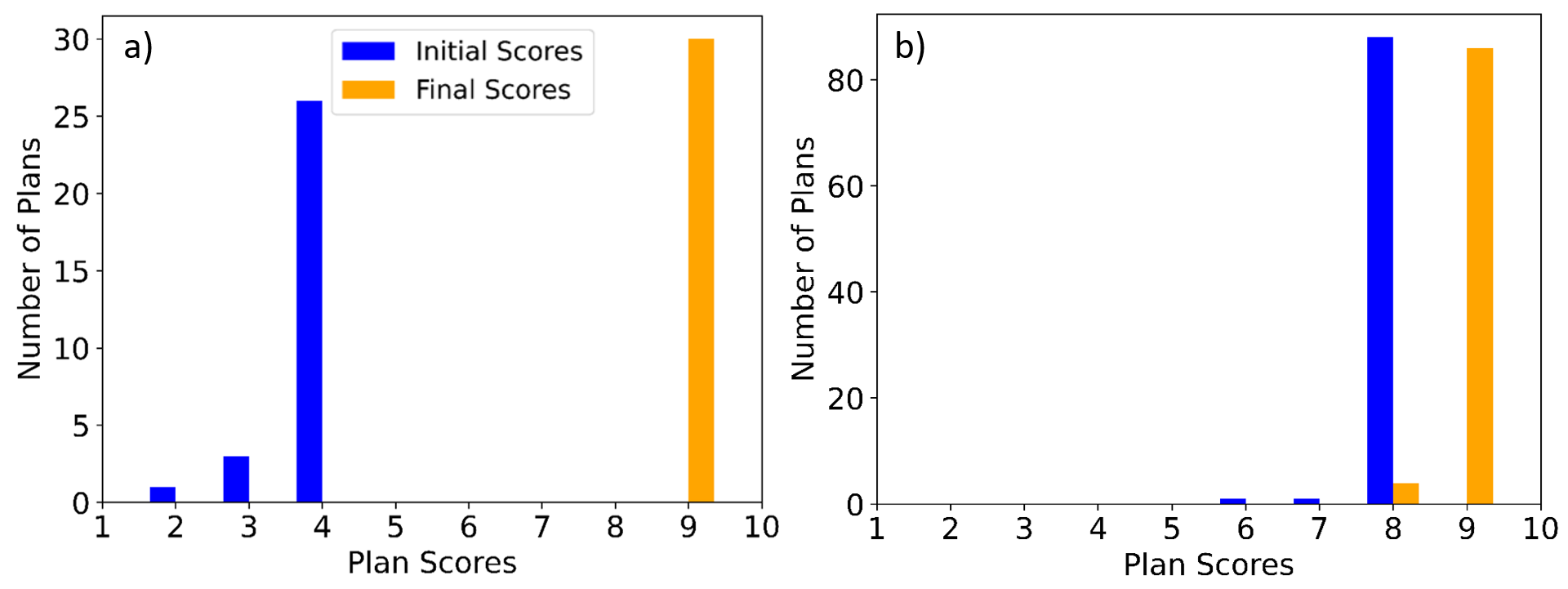} 
    \caption{The plan score distributions before and after actor critic with experience replay (ACER)-guided treatment planning for (a) 30 test cases generated from a single patient case in dataset 2, and (b) 90 test cases generated from 30 patient cases in dataset 3 under random treatment planning parameter (TPP) initializations.\\}
    \label{fig:datasetRandom&2&3}
\end{figure}

We then generate 147 treatment plans by random initialization of TPPs for the same 49 patients (3 random plans for each patient) and perform ACER-guided treatment planning. The results are shown in Figure \ref{fig:dataset1} (b) and (d). In Figure \ref{fig:dataset1} (b), the patient cases are grouped into 7 categories based on their plan scores, using the same method as in Figure \ref{fig:dataset1} (a). Before ACER-guided treatment planning, the mean and std. of the treatment plans are $6.33 \pm 1.65$. 135 out of 147 cases achieve a full score of 9, and 12 cases reach a score of 8, with no cases scoring below 8. The corresponding mean and std. after planning are $8.92 \pm 0.27$. The corresponding patient group distribution is shown in Figure \ref{fig:dataset1} (d), which also shows that ACER-guided treatment planning improve the plan score uniformly across all patient groups, demonstrating the efficacy of this planning agent.

\begin{figure}[h!] 
    \centering
    \includegraphics[width=0.8\textwidth]{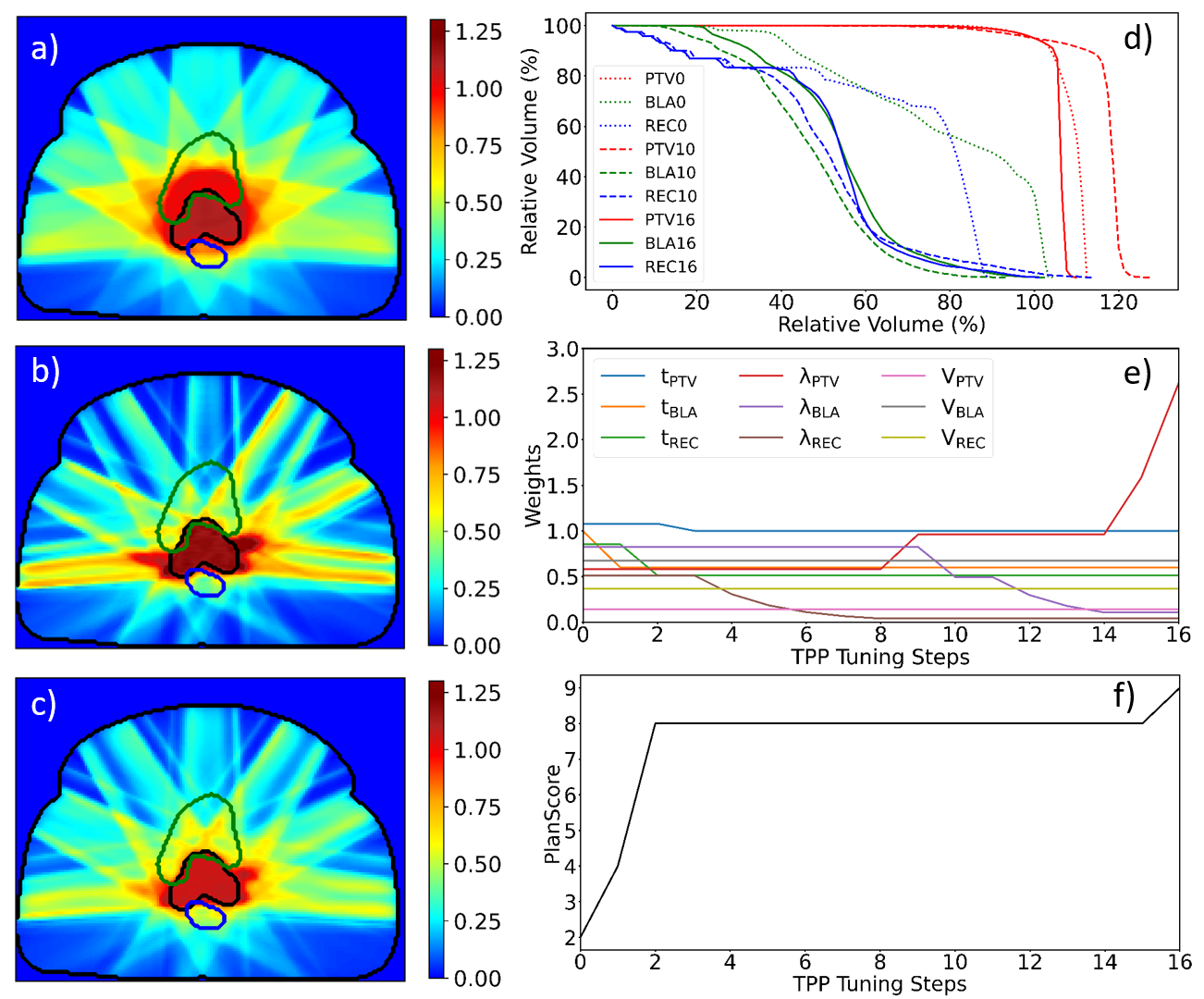} 
    \caption {(a-c) The dose colorwash for a representative test case in step 0, step 10, and final step 16, respectively, under actor critic with experience replay (ACER)-based automatic treatment planning. The contours are represented in black for the prostate target, blue for the rectum, and green for the bladder. In the color bar, color '1' corresponds to the prescription dose level. (d) The corresponding dose volume histograms (DVHs) for steps in (a)-(c). (e) The treatment planning parameter (tpp) tuning choices made by ACER for each planning step. (f) The corresponding plan scores over the 16 steps (the maximum score is 9).\\}
    \label{fig:CORTSample}
\end{figure}

We then extend the network test to datasets 2 and 3 to assess the generality of the trained network on patient cases with features distinct from the training and validation cases. Considering dataset 2 contains only one patient case, we enrich the test by randomly initializing the TPPs 30 times to create 30 independent initial plans. As for dataset 3, it contains 30 different patient cases. Yet, the PTV and OAR data are in a sampled format. We randomly initialize the TPPs for 3 times for each patient case to generate 90 initial treatment plans. For both datasets, the ACER-guided treatment planning is applied to optimize these treatment plans. The results are shown in Figure \ref{fig:datasetRandom&2&3} (a) and (b). From \ref{fig:datasetRandom&2&3}(a), the plan scores for the initial plans in dataset 2 have a mean $\pm$ std. of $3.91 \pm 0.26$. After ACER-based treatment planning, all 30 cases have been elevated to 9. Figure \ref{fig:datasetRandom&2&3} (b) shows that for dataset 3, the initial plans have a mean $\pm$ std. of $7.98 \pm 0.133$. After ACER-guided treatment planning, 86 out of the 90 cases have been improved to a score of 9. The corresponding mean $\pm$ std. after planning is $8.96 \pm 0.21$. These results further demonstrate the efficacy of ACER agent in high quality treatment planning across different datasets.


Combing all test cases, the mean $\pm$ std. of the plan score distributions before ACER-based treatment planning is $6.20 \pm 1.84$. After implementing ACER-based treatment planning, $93.09\%$ of the cases achieve a perfect score of 9, with only $6.12\%$ scoring between 8 and 9, $0.78\%$ scoring between 7 and 8, and no cases scoring below 7. The mean $\pm$ std. of the final scores is $8.93 \pm 0.27$.

Furthermore, we illustrate how ACER-based VTP observes an intermediate treatment plan and makes the TPP adjustment decision for a representative testing case in Figure \ref{fig:CORTSample}. As is shown in Figure \ref{fig:CORTSample} (a) and (d), at the initial step, the plan fails to spare the bladder, partially fails to spare the rectum, and has a hotspot in the PTV, resulting in a low initial plan score of 2. The VTP observes this plan and decides to lower the threshold dose value for the bladder ($t_\mathrm{BLA}$) in the first step, which improves the plan score to 4 by fully sparing the bladder volume. It then continues to enhance rectum sparing by lowering the threshold value for the rectum ($t_\mathrm{REC}$), However, these adjustments result in an even hotter PTV. To address this issue, over the next 14 steps, the ACER-based VTP reduces the priorities for the OARs ($\lambda_\mathrm{REC}$ and $\lambda_\mathrm{BLA}$), lowers threshold dose value in PTV ($t_\mathrm{PTV}$), and increases the PTV priority ($\lambda_\mathrm{PTV}$) until reaching a score of 9. These actions relax OAR constraints while tightening PTV constraints, mirroring the adjustments a human dosimetrist would make. This indicates that the ACER-based agent exhibits a human-like approach to TPP tuning.

\begin{figure}[ht] 
    \centering
    \includegraphics[width=0.75\textwidth]{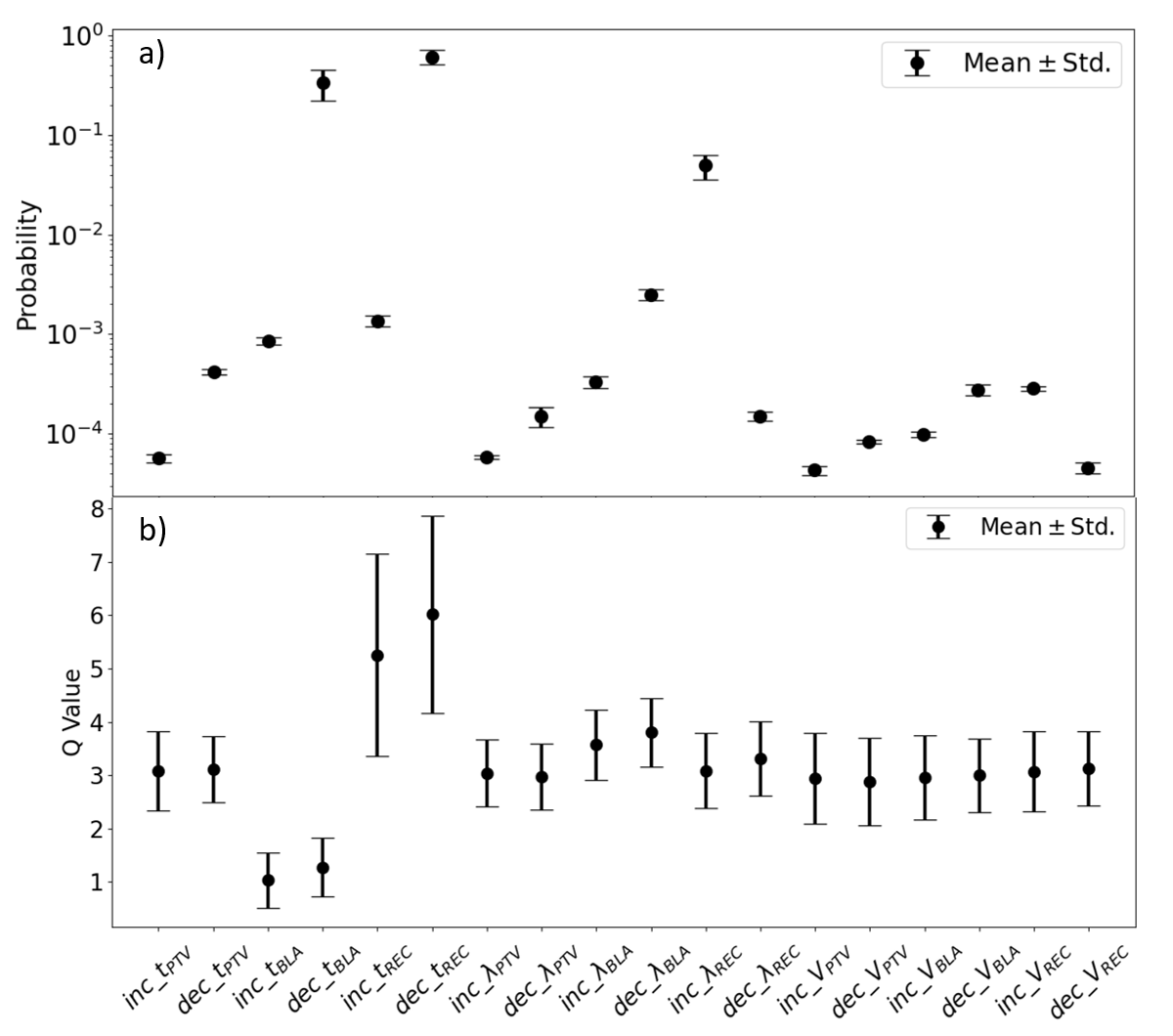} 
    \caption{Policy distributions from the actor critic with experience replay (ACER) agent (top row) and Q value distributions from the deep Q network (DQN) agent \cite{sprouts2022development} (bottom row) for 5 patient cases with similar plan qualities. In all patient cases, the plans fail to spare the rectum with losing all 4 credits for the 4 rectum dose-volume criteria as shown in Table \ref{table:TPPtuning}.\\}
    \label{fig:ProbAction}
\end{figure}

Finally, since the ACER-based treatment planning agent utilizes a stochastic policy, it is important to understand its policy behavior to ensure stability in guiding treatment planning. To investigate this, we identify 5 common cases from ACER-guided and DQN-guided treatment plannings, each with an initial plan score of 5 due to the failure in rectum dose sparing. The mean and standard deviation of the policy distributions and Q-value distributions for the leading 18 actions in these cases are shown in Figure \ref{fig:ProbAction} (a) and (b), respectively. As expected, both networks exhibit relatively stable preferences for certain actions when faced with similar input plans. However, it is noteworthy that ACER strongly prioritizes the reasonable action for rectum sparing, that is to 'decrease $t_\mathrm{REC}$', with a magnitude of order higher than other actions. It also significantly suppresses non-reasonable actions like 'increase $t_\mathrm{REC}$' or 'decrease $\lambda_\mathrm{REC}$', which are several orders of magnitude lower than the leading action. The sub-leading actions, such as 'decrease $t_\mathrm{BLA}$' or 'increase $\lambda_\mathrm{REC}$', are either reasonable or have unclear effects. This type of policy distribution allows the agent to explore the action space while maintaining effectiveness. In contrast, DQN does not significantly differentiate between reasonable and non-reasonable actions. In fact, the leading two actions in DQN have contradictory effects on rectum dose sparing. This further highlights the superior performance of the ACER-based VTP agent. 

\subsection{Adversarial Attack}
\begin{figure}[!ht] 
    \centering
    \includegraphics[width=1\textwidth]{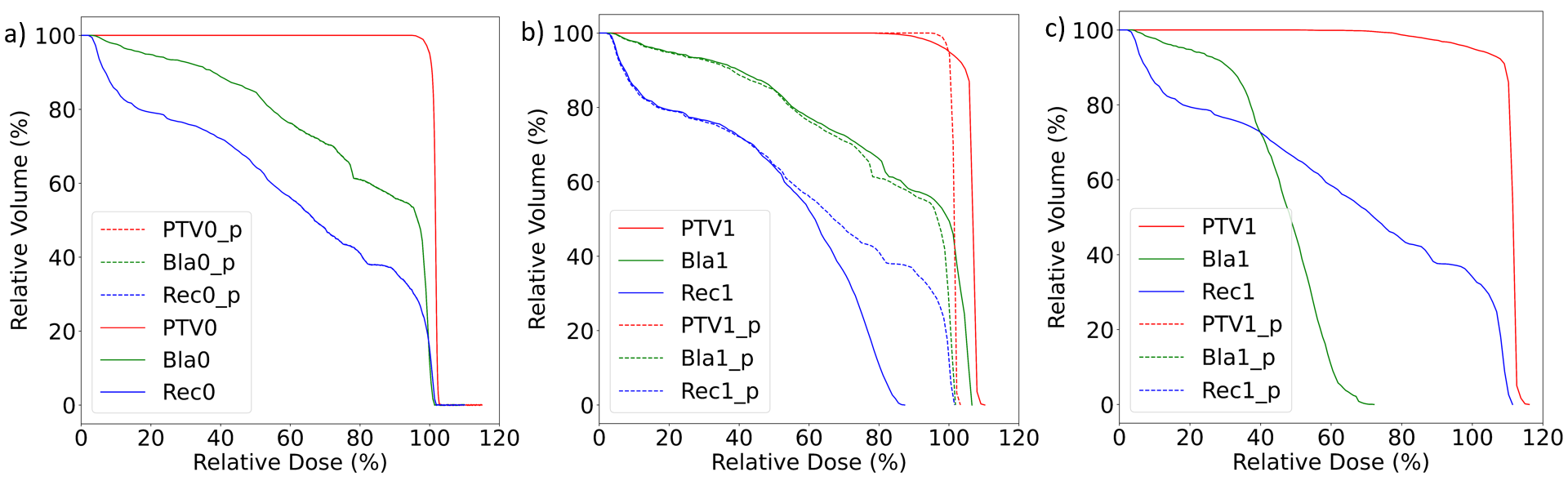} 
    \caption{The illustration of the attack effect from fast gradient sign method (FGSM) on the deep Q network (DQN) \cite{sprouts2022development} and the actor critic with experience replay (ACER) agent based treatment planning. (a) The dose volume histogram (DVH) distributions before and after the perturbation with $\epsilon = 0.001$. The corresponding DVH distributions generated by the in-house treatment planning system (TPS) under the tuned treatment planning parameters (TPPs) by DQN (b) and ACER-based agent (c) before and after the perturbation.\\}
    \label{fig:attack}
\end{figure}

We randomly choose 30 treatment plans and apply one FGSM attack under each value of $\epsilon$ for both ACER and DQN networks. With three $\epsilon$ values of 0.001, 0.01, and 0.1, a total of 180 attacks are performed. The results are shown in Figure \ref{fig:attack} and Table \ref{table:attack}.

An illustration of the attack effect with perturbation to the input states at the $\epsilon=0.001$ level is shown in Figure \ref{fig:attack}. Figure \ref{fig:attack} (a) shows the DVH distributions before and after the perturbation with $\epsilon = 0.001$, which is not distinguishable by naked eyes. The initial plan has a score of 3, which partially fails to spare the rectum and the bladder (both lose 3 points following criteria in Table \ref{table:TPPtuning}). Before the perturbation is applied, DQN produces a $Q$-value distribution that maximizes the action 'decrease $t_\mathrm{REC}$', which improves the plan score to 4. However, after the perturbation, the Q value for the initial top action is reduced by $16.52\%$, and the optimal action changes to "increase $\lambda_\mathrm{REC}$", slightly decreasing the plan score to 2.26. The treatment plans generated under original action and perturbation changed action are shown in Figure \ref{fig:attack} (b). In contrast, for the same patient case, ACER’s policy distribution prioritizes "decreasing $t_\mathrm{BLA}$" and "decrease $t_\mathrm{REC}$" as the top two leading actions, with probabilities of 0.86 and 0.19, respectively (the sum of all action probabilities is 1). The ACER-based stochastic policy selects 'decrease $t_\mathrm{BLA}$', which improves the plan score to 5. After the perturbation is applied, the policy distribution retains the same ranking of actions, with the probabilities of the leading two actions changing by $-0.05\%$ and $0.35 \%$, respectively. The new treatment plans generated are shown in Figure \ref{fig:attack} (c). This demonstrates ACER's stable performance under adversarial attack.

The statistical results for all 180 attacks are shown in Table \ref{table:attack}. As the perturbation level increases from $\epsilon = 0.001$ to $\epsilon = 0.1$, the attack success rate increases from $0.0\%$ to $53.3\%$ for the ACER agent, while it rises from $30.0\%$ to $93.3\%$ for the DQN agent. This indicates that the ACER network demonstrates greater robustness compared to the DQN. To further understand the effects of the attacks, we analyzed changes in action probabilities for ACER and Q values for DQN under various attack levels. We compared these with the differences in probabilities and Q values between the top two leading actions. The analysis reveals that the probability difference between the top two leading actions averages at $67.08\% \pm 21.30\%$ for the ACER agent. Comparing to it, the mean probability change for the leading action in ACER is negligible at $\epsilon = 0.001$ and 0.01 levels. In contrast, for the DQN agent, the mean Q value changes at all perturbation levels are significantly greater than the Q value difference between the top two leading actions. This behavior further explains the relative robustness of the ACER agent against FGSM attacks.\\[0.25 cm]

\begin{table}[ht]
    \caption{The statistical performance of actor critic with experience replay (ACER) and deep Q network (DQN) based agents under three levels of adversarial attacks.}
    \label{table:attack}
    \centering
    \renewcommand{\arraystretch}{1.3}
    \begin{tabularx}{\textwidth}{c*{4}{Y}}
        \hline
        {}&{}&ACER&DQN\\ \hline
        \multirow{2}{*}{$\epsilon =0.001$} & success rate & $0.0\%$&$30.0\%$\\
        &$\Delta P_{1}$ &$-0.194\% \pm 0.25\%$&$-18.55\% \pm 7.73\%$\\ \hline
        \multirow{2}{*}{$\epsilon =0.01$} & success rate & $3.3\%$&$100\%$\\
        &$\Delta P_{1}$ &$0.08 \%\pm 10.24\%$&$-87.95\% \pm 17.98\%$\\ \hline
        \multirow{2}{*}{$\epsilon =0.1$} & success rate & $53.3\%$&$93.3\%$\\
        &$\Delta P_{1}$ &$-36.81\% \pm 46.49\%$&$-68.99\% \pm 68.94\%$\\ \hline  
        {$\Delta P_{12}$} & {} & $67.08\% \pm 21.30\%$&$7.19\% \pm 5.19\%$\\ \hline         
    \end{tabularx}
\end{table}

\section{Discussion}
In summary, we have developed an ACER-based VTP agent that can effectively guide the in-house TPS for inverse treatment planning with prostate IMRT patient cases as testbed. Despite being trained on a single patient case with random initializations of TPPs, the network has demonstrated superior test performance compared to our previously developed DQN-based VTP agent \cite{sprouts2022development}. It has also demonstrated strong generalization, performing well on data from distinct resources, and has exhibited stable performance under adversarial attacks.  This raises the question: why does ACER perform so much better?

First, ACER is stochastic policy based, allowing it to explore a wider solution space without suffering from scaling issues typically associated with DQN. The application of the entropy regularization term during the policy update promotes global convergence. Additionally, ACER incorporates various strategies for effectively managing bias and variance during policy updates. These attributes likely contribute to the network's improved performance in treatment planning. Further investigations are needed to explore how these features impact the network's efficacy.

Second, the greater robustness of the ACER agent compared to the DQN agent in the FGSM attack can be understood from two perspectives. First, the stochastic policy strategy of ACER contributes to better convergence compared to DQN. This aligns with findings that show A3C outperforms DQN in adversarial attacks on Atari games \cite{huang2017adversarial}. Second, unlike tasks such as Atari, which typically have only one optimal action, inverse treatment planning often involves multiple reasonable TPP tuning actions that can enhance plan quality. A well-trained ACER network that effectively prioritizes these reasonable TPP tuning strategies over less suitable options can contribute to its stable performance, even under adversarial attacks. 

In this study, we utilize an in-house TPS for inverse treatment planning during ACER agent training, which may raise concerns about the ACER agent's ability to operate a commercial TPS for high-quality treatment planning. However, we have previously demonstrated the high efficacy of a DRL agent trained with our in-house TPS in operating commercial TPS to achieve high-quality treatment planning \cite{sprouts2022development}. This suggests that the current ACER-based agent, trained with the same in-house TPS, can also function effectively with commercial TPS. We plan to evaluate the performance of the ACER agent on commercial TPS in future work.

Additionally, we use a relatively simple plan quality evaluation system in developing the ACER agent, which may result in differences in plan quality compared to those assessed under clinical evaluation systems. However, we want to emphasize that the primary goal of this work is to demonstrate the efficacy of the ACER agent in treatment plan parameter tuning under a specific reward system. We do not anticipate convergence issues under different evaluation systems.

Another limitation is that we utilize the computationally-efficient FGSM adversarial attack to test the robustness of ACER-based VTP agent. Although this method has shown effectiveness in fooling A3C network on other tasks \cite{huang2017adversarial}, it may not be effective in attacking this VTP agent where multiple suitable actions exist. In our future work, we will test the robustness of this VTP agent by employing stronger adversarial attacks. 

Finally, in this initial study, we employ the ACER network with a discretized action space, consistent with our previous DQN-based development efforts. However, to make this tool practical for treatment planning in real clinical settings, it is essential to train a DRL agent capable of tuning each TPP in a continuous space, and ideally, tuning multiple TPPs in a single planning step. Based on the results of this study, we have identified the potential of ACER to produce a prioritized TPP tuning space that could facilitate multi-TPP tuning in one step. To advance this approach, we may need to further optimize the TPP tuning space to prioritize reasonable TPP strategies effectively. This could be achieved by enhancing the entropy regularization term, activating the TRPO strategy, and enriching the training datasets. Additionally, ACER has a continuous-action counterpart that can be used to explore for continuous TPP tuning. These will be our next step work in the future.

This also raises another important issue worth exploring: the TPP tuning hyperspace. Since the beginning of inverse treatment planning, this hyperspace has remained largely unexplored, with dosimetrists relying on intuition and experience to navigate it. A DRL-based VTP could effectively map this hyperspace and generate insights that dosimetrists can learn from, as illustrated by the prioritized actions derived from the ACER policy distribution. We believe that further investigation into the continuous and simultaneous tuning of multiple TPPs using advanced DRL strategies can not only enhance automation in real-time treatment planning but also contribute to a deeper understanding of the TPP tuning hyperspace. Much like how DRL algorithms have revolutionized game strategies, like in chess, a well-designed DRL agent has the potential to reshape conventional approaches to TPP tuning, ultimately guiding the radiotherapy clinics toward more effective treatment planning.

\section{Conclusion}
We have trained a deep reinforcement network with actor-critic experience replay technique for automatic treatment planning. The trained network can guide the in-house treatment planning system for high quality treatment planning when using the prostate cancer IMRT as a testbed. The trained network has high generality, which performs well over patient data from sources distinct from the training patient dataset. It also shows high robustness over adversarial attack, demonstrating its potential in practical treatment planning in clinical settings.
\section*{Acknowledgment}
This work is partially supported by the rising stars program of UT system and national institutes of health (NIH)/national cancer institute (NCI) grants R37CA214639, R01CA254377, and R01CA237269.

\section*{References}
\bibliography{references}

\begin{thebibliography}{10}

\bibitem{webb2003physical}
S.~Webb,
\newblock The physical basis of IMRT and inverse planning,
\newblock The British journal of radiology {\bf 76}, 678--689 (2003).

\bibitem{li2020automatic}
X.~Li, J.~Zhang, Y.~Sheng, Y.~Chang, F.-F. Yin, Y.~Ge, Q.~J. Wu, and C.~Wang,
\newblock Automatic IMRT planning via static field fluence prediction (AIP-SFFP): a deep learning algorithm for real-time prostate treatment planning,
\newblock Physics in Medicine \& Biology {\bf 65}, 175014 (2020).

\bibitem{7780459}
K.~He, X.~Zhang, S.~Ren, and J.~Sun,
\newblock Deep Residual Learning for Image Recognition,
\newblock in {\em 2016 IEEE Conference on Computer Vision and Pattern Recognition (CVPR)}, pages 770--778, 2016.

\bibitem{8099726}
G.~Huang, Z.~Liu, L.~Van Der~Maaten, and K.~Q. Weinberger,
\newblock Densely Connected Convolutional Networks,
\newblock in {\em 2017 IEEE Conference on Computer Vision and Pattern Recognition (CVPR)}, pages 2261--2269, 2017.

\bibitem{vandewinckele2022treatment}
L.~Vandewinckele, S.~Willems, M.~Lambrecht, P.~Berkovic, F.~Maes, and W.~Crijns,
\newblock Treatment plan prediction for lung IMRT using deep learning based fluence map generation,
\newblock Physica Medica {\bf 99}, 44--54 (2022).

\bibitem{lempart2021volumetric}
M.~Lempart, H.~Benedek, C.~J. Gustafsson, M.~Nilsson, N.~Eliasson, S.~B{\"a}ck, P.~M. af~Rosensch{\"o}ld, and L.~E. Olsson,
\newblock Volumetric modulated arc therapy dose prediction and deliverable treatment plan generation for prostate cancer patients using a densely connected deep learning model,
\newblock Physics and imaging in radiation oncology {\bf 19}, 112--119 (2021).

\bibitem{ronneberger2015u}
O.~Ronneberger, P.~Fischer, and T.~Brox,
\newblock U-net: Convolutional networks for biomedical image segmentation,
\newblock in {\em Medical image computing and computer-assisted intervention--MICCAI 2015: 18th international conference, Munich, Germany, October 5-9, 2015, proceedings, part III 18}, pages 234--241, Springer, 2015.

\bibitem{kandalan2020dose}
R.~N. Kandalan, D.~Nguyen, N.~H. Rezaeian, A.~M. Barrag{\'a}n-Montero, S.~Breedveld, K.~Namuduri, S.~Jiang, and M.-H. Lin,
\newblock Dose prediction with deep learning for prostate cancer radiation therapy: model adaptation to different treatment planning practices,
\newblock Radiotherapy and Oncology {\bf 153}, 228--235 (2020).

\bibitem{finlayson2019adversarial}
S.~G. Finlayson, J.~D. Bowers, J.~Ito, J.~L. Zittrain, A.~L. Beam, and I.~S. Kohane,
\newblock Adversarial attacks on medical machine learning,
\newblock Science {\bf 363}, 1287--1289 (2019).

\bibitem{sutton2018reinforcement}
R.~S. Sutton and A.~G. Barto,
\newblock {\em Reinforcement learning: An introduction},
\newblock MIT press, 2018.

\bibitem{mnih2015human}
V.~Mnih et~al.,
\newblock Human-level control through deep reinforcement learning,
\newblock nature {\bf 518}, 529--533 (2015).

\bibitem{shen2019intelligent}
C.~Shen, Y.~Gonzalez, P.~Klages, N.~Qin, H.~Jung, L.~Chen, D.~Nguyen, S.~B. Jiang, and X.~Jia,
\newblock Intelligent inverse treatment planning via deep reinforcement learning, a proof-of-principle study in high dose-rate brachytherapy for cervical cancer,
\newblock Physics in Medicine \& Biology {\bf 64}, 115013 (2019).

\bibitem{shen2020operating}
C.~Shen, D.~Nguyen, L.~Chen, Y.~Gonzalez, R.~McBeth, N.~Qin, S.~B. Jiang, and X.~Jia,
\newblock Operating a treatment planning system using a deep-reinforcement learning-based virtual treatment planner for prostate cancer intensity-modulated radiation therapy treatment planning,
\newblock Medical physics {\bf 47}, 2329--2336 (2020).

\bibitem{shen2021improving}
C.~Shen, L.~Chen, Y.~Gonzalez, and X.~Jia,
\newblock Improving efficiency of training a virtual treatment planner network via knowledge-guided deep reinforcement learning for intelligent automatic treatment planning of radiotherapy,
\newblock Medical physics {\bf 48}, 1909--1920 (2021).

\bibitem{shen2021hierarchical}
C.~Shen, L.~Chen, and X.~Jia,
\newblock A hierarchical deep reinforcement learning framework for intelligent automatic treatment planning of prostate cancer intensity modulated radiation therapy,
\newblock Physics in Medicine \& Biology {\bf 66}, 134002 (2021).

\bibitem{sprouts2022development}
D.~Sprouts, Y.~Gao, C.~Wang, X.~Jia, C.~Shen, and Y.~Chi,
\newblock The development of a deep reinforcement learning network for dose-volume-constrained treatment planning in prostate cancer intensity modulated radiotherapy,
\newblock Biomedical physics \& engineering express {\bf 8}, 045008 (2022).

\bibitem{gao2023implementation}
Y.~Gao, C.~Shen, X.~Jia, and Y.~K. Park,
\newblock Implementation and evaluation of an intelligent automatic treatment planning robot for prostate cancer stereotactic body radiation therapy,
\newblock Radiotherapy and Oncology {\bf 184}, 109685 (2023).

\bibitem{zhu2021overview}
J.~Zhu, F.~Wu, and J.~Zhao,
\newblock An overview of the action space for deep reinforcement learning,
\newblock in {\em Proceedings of the 2021 4th International Conference on Algorithms, Computing and Artificial Intelligence}, pages 1--10, 2021.

\bibitem{huang2017adversarial}
S.~Huang, N.~Papernot, I.~Goodfellow, Y.~Duan, and P.~Abbeel,
\newblock Adversarial attacks on neural network policies,
\newblock arXiv preprint arXiv:1702.02284  (2017).

\bibitem{wang2016sample}
Z.~Wang, V.~Bapst, N.~Heess, V.~Mnih, R.~Munos, K.~Kavukcuoglu, and N.~De~Freitas,
\newblock Sample efficient actor-critic with experience replay,
\newblock arXiv preprint arXiv:1611.01224  (2016).

\bibitem{mnih2016asynchronous}
V.~Mnih, A.~P. Badia, M.~Mirza, A.~Graves, T.~Lillicrap, T.~Harley, D.~Silver, and K.~Kavukcuoglu,
\newblock Asynchronous methods for deep reinforcement learning,
\newblock in {\em International conference on machine learning}, pages 1928--1937, PMLR, 2016.

\bibitem{lin1992self}
L.-J. Lin,
\newblock Self-improving reactive agents based on reinforcement learning, planning and teaching,
\newblock Machine learning {\bf 8}, 293--321 (1992).

\bibitem{varian2014eclipse}
Varian,
\newblock Eclipse Photon and Electron Algorithms Reference Guide,
\newblock (2014).

\bibitem{sutton1999policy}
R.~S. Sutton, D.~McAllester, S.~Singh, and Y.~Mansour,
\newblock Policy gradient methods for reinforcement learning with function approximation,
\newblock Advances in neural information processing systems {\bf 12} (1999).

\bibitem{munos2016safe}
R.~Munos, T.~Stepleton, A.~Harutyunyan, and M.~Bellemare,
\newblock Safe and efficient off-policy reinforcement learning,
\newblock Advances in neural information processing systems {\bf 29} (2016).

\bibitem{schulman2015trust}
J.~Schulman, S.~Levine, P.~Abbeel, M.~Jordan, and P.~Moritz,
\newblock Trust region policy optimization,
\newblock in {\em International conference on machine learning}, pages 1889--1897, PMLR, 2015.

\bibitem{tieleman2012lecture}
T.~Tieleman,
\newblock Lecture 6.5-rmsprop: Divide the gradient by a running average of its recent magnitude,
\newblock COURSERA: Neural networks for machine learning {\bf 4}, 26 (2012).

\bibitem{craft2014shared}
D.~Craft, M.~Bangert, T.~Long, D.~Papp, and J.~Unkelbach,
\newblock Shared data for intensity modulated radiation therapy (IMRT) optimization research: the CORT dataset,
\newblock GigaScience {\bf 3}, 2047--217X (2014).

\bibitem{breedveld2017data}
S.~Breedveld and B.~Heijmen,
\newblock Data for TROTS--the radiotherapy optimisation test set,
\newblock Data in brief {\bf 12}, 143--149 (2017).

\bibitem{szegedy2013intriguing}
C.~Szegedy, W.~Zaremba, I.~Sutskever, J.~Bruna, D.~Erhan, I.~Goodfellow, and R.~Fergus,
\newblock Intriguing properties of neural networks,
\newblock arXiv preprint arXiv:1312.6199  (2013).

\bibitem{biggio2018wild}
B.~Biggio and F.~Roli,
\newblock Wild patterns: Ten years after the rise of adversarial machine learning,
\newblock in {\em Proceedings of the 2018 ACM SIGSAC Conference on Computer and Communications Security}, pages 2154--2156, 2018.

\bibitem{goodfellow2014explaining}
I.~J. Goodfellow, J.~Shlens, and C.~Szegedy,
\newblock Explaining and harnessing adversarial examples,
\newblock arXiv preprint arXiv:1412.6572  (2014).

\end{thebibliography}
\end{document}